\documentstyle[aps,prl,epsf]{revtex}

\begin{document}
\draft
\tighten
\twocolumn[\hsize\textwidth\columnwidth\hsize\csname
@twocolumnfalse\endcsname
\title{Interaction of a quantum dot with an incompressible
two-dimensional electron gas}
\author{V.M. Apalkov and Tapash Chakraborty$^*$}
\address{Max-Planck-Institut f\"ur Physik komplexer Systeme, 
01187 Dresden, Germany}
\date{\today}
\maketitle
\begin{abstract}
We consider a system of a incompressible quantum Hall liquid
in close proximity to a parabolic quantum dot containing a 
few electrons. We observe a significant influence of the 
interacting electrons in the dot on the excitation spectrum 
of the incompressible state in the electron plane. Our calculated 
charge density indicates that unlike in the case of an impurity, 
interacting electrons in the dot seem to confine the fractionally 
charged excitations in the incompressible liquid.
\end{abstract}
\vskip2pc]
\narrowtext
A two-dimensional electron gas under the influence of a strong,
external magnetic field exhibits the celebrated fractional 
quantum Hall effect (FQHE) \cite{qhbook,advances}, that has been 
the subject of intense investigation for almost two decades. 
One of the most successful description of the quantum state in 
such a system is the Laughlin state \cite{laugh}, which describes 
the ground state of an electron system with $\frac13$-filled
lowest Landau level, as a highly correlated incompressible 
liquid. Incompressibility of the state implies the existence of
a gap in the excitation spectrum and that the low-lying 
elementary excitations are fractionally-charged quasiparticles 
and quasiholes \cite{qhbook,advances,laugh}. A large body of 
theoretical and experimental work has already established many 
of these properties of the unique quantum state. Dispersion of 
various collective modes in the FQHE has also been well 
established \cite{advances}.

Quantum Hall (QH) properties of the electron gas, such as the 
ones mentioned above, are however for electron motion in a plane. 
On the other hand, when the electron motion in the plane is 
further quantized in the planar two dimensions, we get what 
is known as a quantum dot (QD) \cite{chakmak,qdbook}. Quantum 
dots represent the ultimate reduction in the dimensionality of 
a semiconductor device where electrons have no kinetic energy
and as a consequence, they have sharp energy levels like in
atoms. These zero-dimensional electron systems have enjoyed
enormous popularity because of their importance in understanding
fundamental concepts of nanostructures and at the same time, for 
their application potentials. Development of extremely small
self-assembled quantum dots (only a few nanometers across) 
containing only a few electrons that can be inserted into the 
dot in a controlled manner, have led to important new device 
applications. In this letter, we propose a system where a 
parabolic QD \cite{chakmak} containing only a few electrons 
is brought in close proximity to a two-dimensional electron gas 
(2DEG) that was initially in the Laughlin state with $\frac13$ 
filling of the lowest Landau level. We investigate the influence
of interacting electrons in the dot on the Laughlin state that
exists in the absence of the quantum dot. 

Why (and how) should one bring a quantum dot in the vicinity of 
a 2DEG? The answer is, a device similar to the type of system 
described here already exists as a single-photon detector 
\cite{shields}. In this device, a layer of nanometer-sized InAs 
quantum dots is placed near a 2DEG, separated by a thin barrier. 
Charge carriers photoexcited by incident light (visible or 
near-infrared wavelengths) are trapped by the dots which results 
in a depletion of the electron density adjacent to the dots. 
As a consequence, the conductivity of the 2DEG is altered.
This change in conductivity is in fact, measurable. Observation 
of step-like rise of conductivity, each step is due to discharge 
of a QD by a single photon, has been associated with the detection 
of a single photoexcited carrier in a single dot \cite{shields}. 
We propose here that, placed in a strong magnetic field, 
incident light on  such a device would perhaps light up the 
quasiparticles of the incompressible state.

Formally, our system is similar to a double-layer FQHE system 
\cite{double,jim}, except that in one of the two layers electrons 
are confined by a harmonic potential 
$$V_{\rm conf}(x,y)=\frac12m^*\omega_0^2\left(x^2+y^2\right),$$
where $\omega_0$ is the confinement potential strength and
the corresponding oscillator length is $l_{\rm dot}=\left(
\hbar/m^*\omega_0\right)^{\frac12}$. We consider Coulomb 
interaction between the electrons in the dot and in the plane. 
In our calculations presented below, the 2DEG is kept at the 
filling factor $\nu=\frac13$ (Laughlin state) and the QD is 
filled with $N_D=1$ or 2 interacting electrons. 

We evaluate the Laughlin state numerically in a spherical
surface geometry which is a well-established method to describe 
the ground state and low-lying excitations at $\frac13$ filling 
factor \cite{haldane}. This geometry is more appropriate in this 
case due to the circular symmetry of the problem. In this case the  
single particle wave function of the electron in the layer has 
the form:
\begin{eqnarray*}
\varphi_m =\left[\frac{2S+1}{4\pi} 
\left(\begin{array}{c} 2S \\ S+m \end{array}\right)\right]^{1/2} 
u^{S+m} v^{S-m} 
\end{eqnarray*}
where $m=-S,\ldots, S$ is $z$-component of the angular momentum of
the electron and $2S$ is the number of flux quanta throughout the sphere
in units of the elementary flux quantum; $u=\cos(\theta/2) e^{i\phi/2}$,
$v=\sin(\theta/2) e^{-i\phi/2}$ and $\theta$, $\phi$ are polar angles;
$(\begin{array}{c} 2S \\ S+m \end{array})$ is the binomial coefficient.

The single particle wave functions in the quantum dot have the usual
form:
\begin{eqnarray*}
\psi_{n,l}(x,\phi) &= &
  \left(\frac{b}{2\pi l_0^2}\frac{n!}{(n+|l|)!}\right)^{1/2} \\
& & \times\sum_{j=0}^n C(n,l,j) e^{-il\phi} e^{-(x^2/2)}
 x^{2j+|l|}
\end{eqnarray*}
where $x=(b/2l_0)^{\frac12}r$, $b=(1+4\omega_0^2/\omega_c^2)^{\frac12}$
and
$$C(n,l,j)=(-1)^j\frac{(n+|l|)!}{(n-j)!(|l|+j)!j!},$$
$n=0,1..$ is the radial quantum number and $l$ is the azimuthal 
quantum number. In our case of a large quantum dot (15 nm) only 
the single particle states with $n=0$ and $l=0,1,..$ are important.
Then the  single particle energy spectrum has the one-dimensional 
oscillator form: 
$l \hbar [(\omega_c^2+\omega_0^2)^{1/2}-\omega_c]/2$.

We also study the electron density distribution for electrons 
in the dot ($\rho_D$) and for the electrons in the layer ($\rho_L$):
\begin{eqnarray*}
\rho_D (r) & = &\int\cdots\int d\vec{r}_{D,1}\ldots d\vec{r}_{L,1}
\ldots \sum_{i=1}^{N_D} \delta (\vec{r}-\vec{r}_{D,i})   \\
 &  & \times \left| \Psi_M (\vec{r}_{D,1},\ldots | \vec{r}_{L,1} \ldots)
    \right|^2
\end{eqnarray*}
\begin{eqnarray*}
\rho_L(r) & = &\int \cdots \int d\vec{r}_{D,1}\ldots
d\vec{r}_{L,1} \ldots \sum_{i=1}^{N_L} 
\delta (\vec{r}-\vec{r}_{L,i})   \\
&  & \times\left|\Psi_M (\vec{r}_{D,1},\ldots |\vec{r}_{L,1}
\ldots)\right|^2
\end{eqnarray*}
where $N_D$ and $N_L$ are the numbers of the electrons in the dot
and in the layer, respectively. The integration over $\vec{r}_{L,i}$ is 
restricted to the sphere. 

The interaction part of the Hamiltonian is
given by the expression
\begin{eqnarray*}
H_{\rm int} &=&\frac{e^2}\varepsilon\sum_{i<j}\frac1{|\vec{r}_{D,j}-
\vec{r}_{D,i}|}+\frac{e^2}\varepsilon\sum_{i<j}\frac1{|\vec{r}_{L,j}-
\vec{r}_{L,i}|} \\
 & &  + \frac{e^2}\varepsilon\sum_{i,j}\frac1{\left[d^2+|\vec{r}_{D,j} -
 \vec{r}_{L,i}|^2\right]^{\frac12}}
 \end{eqnarray*}
where $d$ is the separation between quantum dot and the layer,
$\vec{r}_{D}=(r_{D}\cos\phi_{D},r_{D}\sin\phi_{D})$ is
the two-dimensional vector corresponding to the electron in the 
quantum dot, $\vec{r}_L=(2R\sin(\theta /2)\cos\phi_L, 
2R\sin(\theta/2)\sin\phi_L)$ is the two-dimensional vector 
corresponding to the electron in the sphere with sphere radius 
$R=S^{\frac12}l_0$ and polar angle $\theta$.    

All computations are done for six electrons 
in the layer (sphere) which form the incompressible liquid with 
filling factor $\nu=\frac13$. In this case the sphere radius is 
$R=\sqrt{7.5}l_0$. For electrons in the quantum dot we take  10 lowest 
single particle states. Under such conditions we can consider only 
one and two electrons in the dot. Any additional number of electrons 
in the dot (three and more) requires a larger sphere for electrons 
in the layer which results in a much larger matrix that has to be 
diagonalized numerically. All electrons are treated as 
spinless particles. In what follows we use the magnetic length 
$l_0$ as the unit of length and the Coulomb energy 
${\cal E}_c = e^2/\varepsilon l_0 $ as the unit of energy, where
$\varepsilon$ is the background dielectric constant. In all our 
calculations presented here, the magnetic length is
taken to be $6.6$ nm, which corresponds to the magnetic field of
15 T. For the quantum dots we consider parameters
appropriate to GaAs and the dot size, $l_{\rm dot}=15$ nm.
The size of the dot is dictated by the fact that for smaller
dots, the difference energy (energy difference between the
ground state and the lowest excited state) is much larger
than the lowest energy excitations of the incompressible state
and therefore has no noticeable effect of the spectrum.
The interlayer separation $d$, and as a result, the interlayer 
interaction \cite{double} has also been varied in our 
calculations. Here we consider $d=1.5, 2.0l_0$ separations.
The latter separation was found to be optimum in the double-layer
FQHE systems \cite{double}. Smaller separations tend to 
close the energy gap, the hallmark of the incompressible
state.

In the absence of the quantum dot, states in the spherical geometry
appear as multiplets characterized by the rotational quantum number 
$L$. However, if suppose we place an impurity near the north pole of 
the sphere, then states can be classified only by the azimuthal 
rotational quantum number $M=L_z$, and changes in $M$ indicate charge 
redistribution \cite{impure}. In this geometry, the minima in the 
charge density were identified with the center of a quasiparticle 
defect (fractionally charged) emitted by the impurity. With increasing 
values of $M$, that defect was found to progress outward \cite{impure}. 
Further, due to the incompressibility of the Laughlin state, 
there is no screening of the impurity. Laughlin state was 
found to be stable regardless of the strength of the impurity.

\begin{figure}
\centerline{
\epsfxsize=3.5in
\epsfbox{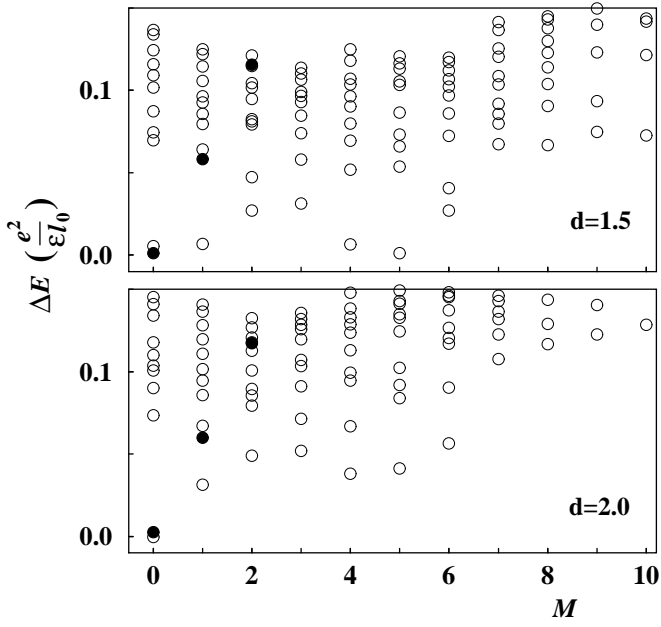}}
\vspace*{0.1in}
\protect\caption[onebody]
{\sloppy{Energy spectrum of a Coulomb-coupled quantum 
dot-quantum Hall system at $\nu=\frac13$ (open circles). 
The quantum dot contains a single electron and is 
separated from the 2DEG by $d=1.5, 2.0l_0$. The filled 
circles are the energies of an isolated QD, presented here 
as a reference.
}}
\label{single}
\end{figure}

\begin{figure}
\centerline{
\epsfxsize=3.2in
\epsfbox{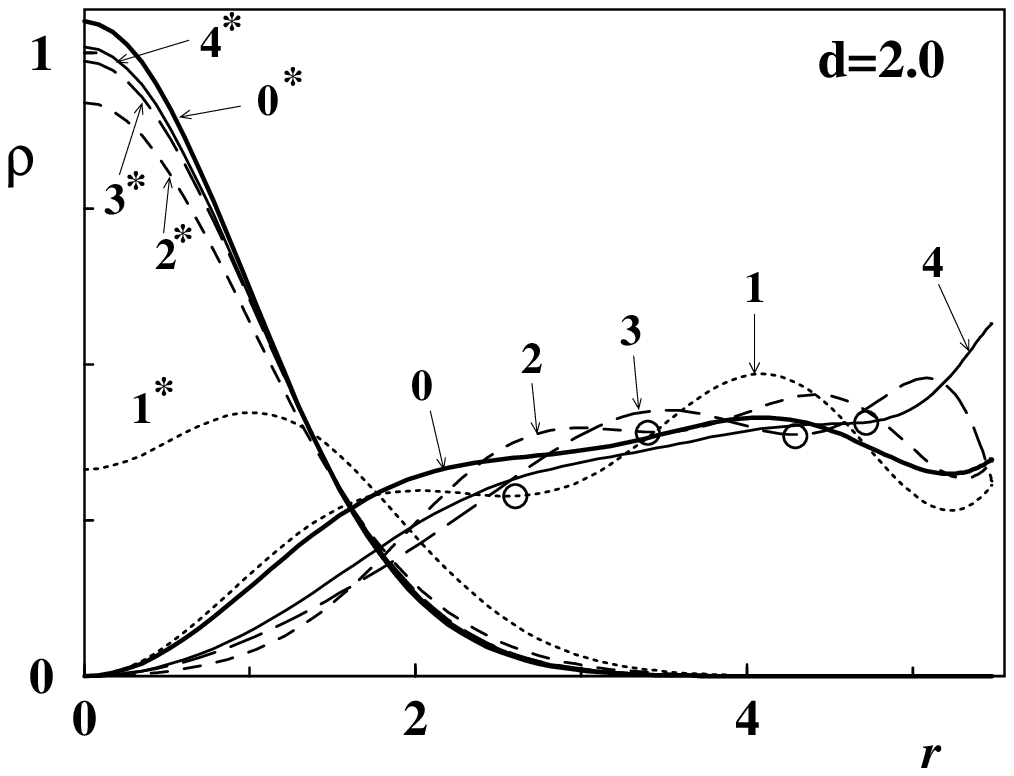}}
\vspace*{0.1in}
\protect\caption[onecharge]
{\sloppy{Charge-density profile $\rho_L(r)$ of the ground 
state and low-lying excitations of a QD+QH system at
$\frac13$ filling factor, corresponding to 
Fig.~\protect\ref{single}, (for $M=0-4$, as indicated in the 
figure) and for the QD, $\rho_D(r)$ with a single
electron (with $M=0^*-4^*)$. Open circles indicate 
the minima in the charge density.
}}
\label{chargeone}
\end{figure}

In Fig.~\ref{single} the energy spectrum of the system with one 
electron in the dot is shown by open circles. This case is closely 
related to the system of incompressible liquid in the field of 
a charged impurity discussed above. However, in our case there 
is an additonal type of collective excitation due to the additional 
degree of freedom of the electron in the dot. For small separation 
($d=1.5$) the perturbation of incompressible liquid by the electron 
in the dot is strong and the collective excitation is 
gapless. For a larger separation ($d=2.0$) there is a well
defined branch of lowest excitations at $M>0$ (Fig.~\ref{chargeone}). 
To understand more about this branch we plot the electron density 
distribution in the layer and in the quantum dot for 
the lowest states with the given $M$ (Fig.~\ref{chargeone}). We 
notice that in the  states with $M=0,2,3,4$ the electron in the 
quantum dot is almost in the ground state of the dot. The excited 
states at $M=2,3,4$ can be described by the process of ionization 
as an emission of the fractionally charged quasihole: the quasihole 
is moving away from the quantum dot with increasing $M$. The positions 
of the quasihole are shown by open circles. This picture is the 
same as for the charged impurity near incompressible liquid 
\cite{impure}. At the same time the state at $M=1$ has different 
nature. It is the collective excitation of the electron in the dot
and the electrons of incompressible liquid. Such low-energy 
excitations can be observed only when the separation between the 
energy levels in the quantum dot ($\hbar[(\omega_c^2+\omega_0^2)^{1/2}
-\omega_c]/2$) is about the incompressible gap ($0.1{\cal E}_c$). 
This type of excitation gives the linear dependance of excitation 
spectra as a function of $M$ for small $M$. 

\begin{figure}
\centerline{
\epsfxsize=3.5in
\epsfbox{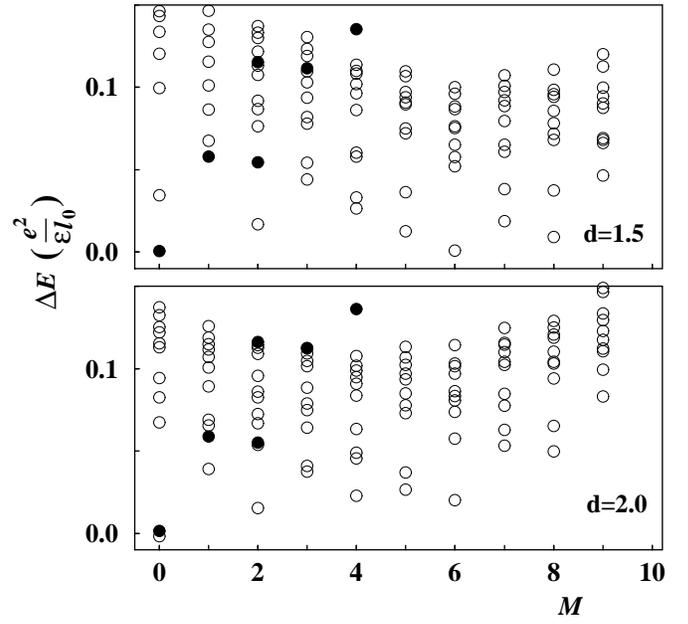}}
\vspace*{0.1in}
\protect\caption[twobody]
{\sloppy{Energy spectrum of a Coulomb-coupled quantum 
dot-quantum Hall system at $\nu=\frac13$ (open circles). 
The quantum dot contains two interacting electrons and is 
separated from the 2DEG by $d=1.5, 2.0l_0$. The filled 
circles are the energies of the isolated QD.  
}}
\label{double}
\end{figure}

\begin{figure}
\centerline{
\epsfxsize=3.2in
\epsfbox{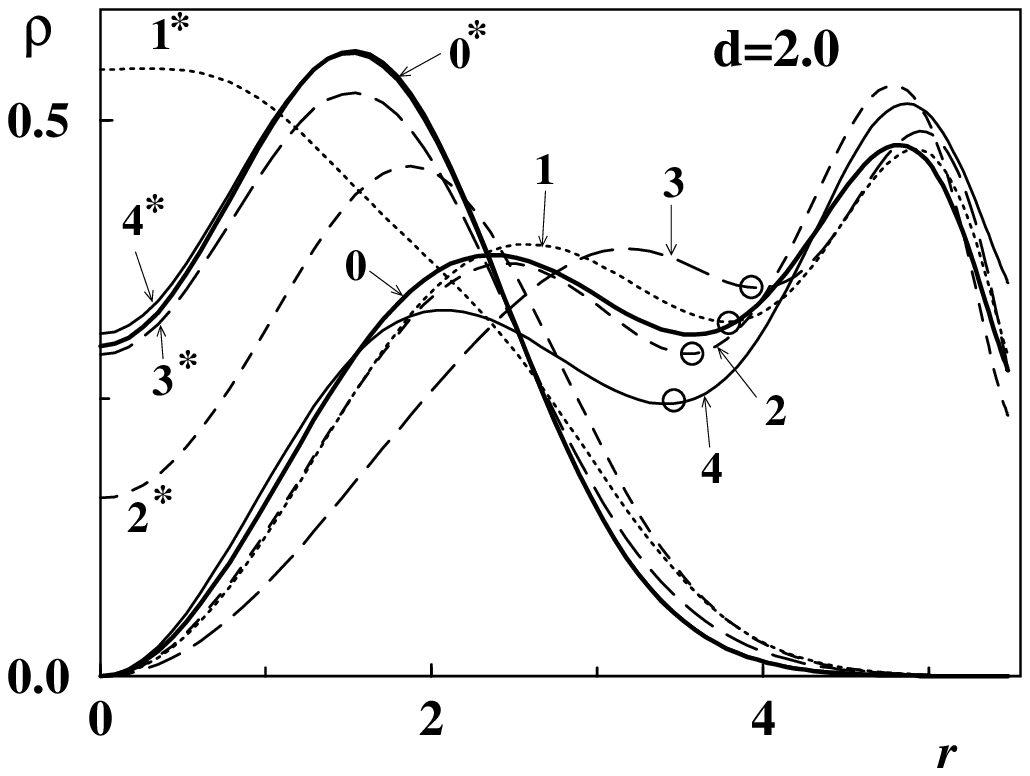}}
\vspace*{0.1in}
\protect\caption[twocharge]
{\sloppy{Charge-density profile $\rho_L(r)$ of the ground 
state and low-lying excitations of a QD+QH system at
$\frac13$ filling factor, corresponding to 
Fig.~\protect\ref{double}, (for $M=0-4$, as indicated in the 
figure) and for the QD, $\rho_D(r)$ with two interacting
electrons (with $M=0^*-4^*)$. Open circles indicate 
the minima in the charge density.
}}
\label{chargetwo}
\end{figure}

A more interesting situation occurs when there are more
than one electron in the dot. In this case the interaction between the 
electrons in the dot makes the quantum dot an impurity center with 
non-trivial charge distribution and with internal degree of freedom.
In Fig.~\ref{double} the energy spectra of the system with 
two electrons in the quantum dot is shown by open circles. The 
states of the pure electron system in the dot are shown by 
filled circles. The angular momentum $M$ is counted from the angular 
momentum of the ground state. For a pure two electron system in 
the quantum dot the angular momentum of the ground state is equal 
to 3. For small separation ($d=1.5$), the incompressible liquid is 
strongly perturbed by the electrons in the dot. The perturbation 
is stronger than that of the single electron case due to the 
larger charge of the dot. At a larger separation ($d=2.0$) the lowest 
branch of the excitation spectra develops an oscillatory structure 
as a function of $M$. The electron density distribution 
(Fig.~\ref{chargetwo}) shows that the state at $M=1$ has different 
distribution compared to the states at $M>1$, and can be 
described in the same manner as for the one electron system, that is
the collective excitation of the electrons in the dot and an
incompressible liquid. The excited states at $M=2,3,4$, however, 
can not be considered simply as the process of ionization. 
In this figure, the position of the minimum of charge distribution, 
shown by open circles exhibit oscillatory behavior with $M$ and
remains almost confined in the same region. These oscillations are
correlated with oscillations in the energy spectra (Fig.~\ref{double}). 
Confinement of the quasihole in the incompressible state by the quantum 
dot is purely due to the interaction between the electrons in the 
quantum dot which results in the specific charge distribution in the 
quantum dot and an additional interaction of a qusihole of the
incompressible liquid with the local excitation of the dot.

In closing, we have explored a system of a quantum dot placed
in close proximity to a two-dimensional electron gas that is in 
the incompressible liquid state. Our results indicate that for 
a single electron in the dot, the physics is somewhat like that 
of an impurity which emits a fractionally-charged quasihole that 
moves away from the dot with increasing $M$. For small $M$, we 
notice a linear behavior of the excitation spectrum. Most importantly, 
however, we find that for two interacting electrons in the quantum 
dot, the collective excitation exhibits an oscillatory behavior 
which is due to confinement of the fractionally-charged quasihole 
excitations by the quantum dot. This is purely a consequence of
interelectron interaction in the dot. Therefore, in a suitable 
set up, the single-photon detector might in fact, be a detector 
for the fractionally-charged excitations of the incompressible 
Laughlin state.

We would like to thank P. Fulde for his support and kind
hospitality in Dresden.


\vfil

\end{document}